\documentclass[12pt]{iopart}
\usepackage{amsfonts,amsmath,amssymb,amsthm}
\usepackage{bm,epsfig,revsymb,upgreek,url}
\usepackage[colorlinks=true,dvipdfm]{hyperref}
\begin{document}

\title[A new criterion for zero quantum discord]{A new criterion for zero quantum discord}

\author{Jie-Hui Huang$^{1,2}$, Lei Wang$^{1,3}$ and Shi-Yao Zhu$^{1,4}$}

\address{$^{1}$Beijing
Computational Science Research Center, Beijing, 100084, China}

\address{$^{2}$Department of Physics, Nanchang
University, Nanchang, 330031, China}

\address{$^{3}$College of Physics,
Jilin University, Changchun 130021, China}

\address{$^{4}$Department of
Physics, Hong Kong Baptist University, Hong Kong, China}

\eads{\mailto{hjhxyx@yahoo.com.cn}}

%\date{\today}

\begin{abstract}
We propose a new criterion to judge zero quantum discord for
arbitrary bipartite states. A bipartite quantum state has zero
quantum discord if and only if all blocks of its density matrix are
normal matrices and commute with each other. Given a bipartite state
with zero quantum discord, how to find out the set of local
projectors, which do not disturb the whole state after being imposed
on one subsystem, is also presented. A class of two-qubit X-state is
used to test the criterion, and an experimental scheme is proposed
to realize it. Consequently, we prove that the positive
operator-valued measurement can not extinguish the quantum
correlation of a bipartite state with nonzero quantum discord.
\end{abstract}


%\submitto{\NJP}

\maketitle

%-----------------------------------------------------------------------
\section{Introduction}
For a bipartite system prepared in an entangled state, a local
measurement on one of the two subsystems will affect the other
subsystem owing to the \lq\lq nonlocal features\rq\rq ~of the
entanglement\cite{1}. However, entanglement is not always necessary
for illustrating the non-localities in a quantum
system\cite{bennett}. In 1998, a model of deterministic quantum
computation with one qubit (DQC1) is proposed for quantum computing
by using highly mixed states\cite{2,2add1}, which has been
experimentally implemented in 2008\cite{3}. It is a good example
illustrating that some highly mixed states, even fully separable,
contain intrinsic quantum correlations, and have potential
applications in the quantum computing. Furthermore, quantum
correlation is found to be more robust than entanglement in a noisy
environment, which makes the quantum algorithms based only on
quantum correlation more robust than those based on
entanglement\cite{4,4add1,4add2}.

If a bipartite quantum state is in a product state,
$\rho=\rho_A\otimes\rho_B$, with $\rho_A$ ($\rho_B$) being the
reduced density matrix of subsystem A (B), the state has no quantum
correlation. However, a state with zero quantum correlation is not
always a product state. The quantum correlation of a bipartite state
is usually measured by quantum discord, introduced by Ollivier and
Zurek in ref.\cite{5}. The question of how to find out whether a
quantum state has zero quantum discord or not is fundamentally
important; it is the first step to distinguish the quantum features
of a bipartite state from the classical. For example, it is shown
that zero quantum discord between a quantum system and its
environment is necessary and sufficient for describing the evolution
of the system through a completely positive map\cite{12,12add1}. In
addition, a quantum state can be locally broadcasted if and only if
it has zero quantum discord \cite{13,13add1}. Recently, a necessary
and sufficient condition for nonzero quantum discord was
proposed\cite{14}, with the help of a correlation matrix, derived
from the density matrix, and its singular value decomposition. In
this paper, we present a simpler method for judging zero quantum
discord, where we only need to partition the density matrix into
$N^2$ block matrices ($N$: the dimension of one subsystem), and
check some properties of these block matrices. This method is valid
for arbitrary bipartite states and easy to implement. An example
with a scheme to experimentally realize it, is proposed in order to
test this criterion. Based on this new criterion, we also prove that
the positive operator-valued measurement (POVM) \cite{8,8add1} can
not extinguish the quantum correlation of a bipartite state with
nonzero quantum discord.

%-----------------------------------------------------------------------
\section{New criterion for zero quantum discord}
Ollivier and Zurek introduced the concept of quantum discord to
quantify the quantum correlation of a bipartite state, which is
defined as the difference between two conditional entropies
(classically equivalent quantities)\cite{5},
\begin{align} \label{equ1}
\delta(\rho_{AB})_{\{|k_B>\}}=H(A|\{|k_B>\})-[H(\rho_{AB})-H(\rho_{B})],
\end{align}
where $H(A|\{|k_B>\})$ is calculated by
$\sum_{k}p_{k_B}H(\rho_{k_B})$ with
$\rho_{k_B}=\frac{1}{p_{k_B}}<k_B|\rho_{AB}|k_B>$ and
$p_{k_B}=\text{Tr}_A(<k_B|\rho_{AB}|k_B>)$, and $H(\rho)$ is the von
Neumann entropy of the quantum state $\rho$\cite{6,7,7add1}. Here
the subsystem A is regarded as the system and B as the apparatus.
$\{|k_B>\}$ represent a set of local projectors on B, rather than
the POVM used in ref.\cite{9}. In the calculation, different set of
projectors will give out different values of quantum discord for the
same quantum state. How to find out the set of local projectors
which yields minimum quantum discord is very
difficult\cite{11,11add1,11add2,11add3,11add4,11add5}.

In a given basis, $\{|i_Ak_B>\}$ ($i=1,2,\cdots,N$ and
$k=1,2,\cdots,M$), arranged as
$\{|1_A1_B>,\cdots,|1_AM_B>,|2_A1_B>,\cdots, |N_AM_B>\}$, an AB
bipartite quantum state can be described by the following density
matrix,
\begin{align} \label{equ2}
%\begin{equation}
\rho_{AB}=\left(
  \begin{array}{ccc}
    \rho_{11} & \cdots & \rho_{1(NM)}\\
    \vdots & \ddots & \vdots\\
    \rho_{(NM)1} & \cdots & \rho_{(NM)(NM)}\\
  \end{array}
\right),
%\end{equation}
\end{align}
which has zero quantum discord if and only if it can also be written
as\cite{5}
\begin{align} \label{equ3}
\rho_{AB}=\sum_{i=1}^N\sum_{j=1}^N\sum_{k^{\prime}=1}^MC_{i_Aj_Ak_B^{\prime}}(|i_A><j_A|)(|k^{\prime}_B><k^{\prime}_B|),
\end{align}
with $C_{i_Aj_Ak_B^{\prime}}$ being real or complex numbers and
$\{|k^{\prime}_B>\}$ ($k^{\prime}=1,2,\cdots,M$) being a particular
set of local projectors on B. The quantum state in the form of
Eq.(\ref{equ3}) is called pointer state\cite{5}, in which one can
locally access the information in the system without changing the
whole density matrix. Since the evaluation of quantum discord is
asymmetric, and depends on which subsystem is chosen as the system
and which one is the apparatus, the zero quantum discord of the
quantum state (\ref{equ3}) with the subsystem B being the apparatus,
does not guarantee the zero quantum discord for A being the
apparatus.

The $(NM)\times(NM)$ matrix in Eq.(\ref{equ2}) can be partitioned
into $N^2$ blocks,
\begin{align} \label{equ4}
%\begin{equation}
\rho_{AB}=\left(
  \begin{array}{ccc}
    \rho^{(1_A1_A)} & \cdots & \rho^{(1_AN_A)}\\
    \vdots & \ddots & \vdots\\
    \rho^{(N_A1_A)} & \cdots & \rho^{(N_AN_A)}\\
  \end{array}
\right)
%\end{equation}
\end{align}
with each block being an $M\times M$ matrix,
\begin{align} \label{equ5}
%\begin{equation}
\rho^{(i_Aj_A)}&=\left(
  \begin{array}{ccc}
    \rho_{((i-1)M+1)((j-1)M+1)} & \cdots & \rho_{((i-1)M+1)(jM)}\\
    \vdots & \ddots & \vdots\\
    \rho_{(iM)((j-1)M+1)} & \cdots & \rho_{(iM)(jM)}\\
  \end{array}
\right),
%\end{equation}
\end{align}
which means the state(\ref{equ2}) or (\ref{equ4}) is equivalent to,
\begin{align} \label{equ6}
%\begin{equation}
\rho_{AB}=\sum_{i=1}^N\sum_{j=1}^N(|i_A><j_A|)\rho^{(i_Aj_A)}.
\end{align}
We rewrite the quantum state Eq.(\ref{equ3}) in the basis
$\{|i_Ak_B>\}$,
\begin{align} \label{equ7}
%\begin{equation}
\rho_{AB}=\sum_{i=1}^N\sum_{j=1}^N\sum_{k=1}^MC_{i_Aj_Ak_B}(|i_A><j_A|)U(|k_B><k_B|)U^{\dag}.
\end{align}
where the local unitary transformation $U$ connects $\{|k_B>\}$ and
$\{|k^{\prime}_B>\}$ through the relation $|k^{\prime}_B>=U|k_B>$.
In order to make Eq.(\ref{equ6}) have the same form of
Eq.(\ref{equ7}), all block matrices $\rho^{(i_Aj_A)}$ must be able
to be diagonalized by the same unitary transformation $U$,
\begin{align} \label{equ8}
\rho^{(i_Aj_A)}=U\left[\sum_{k=1}^MC_{i_Aj_Ak_B}(|k_B><k_B|)\right]U^{\dag},
\end{align}
which gives us the following relation,
\begin{align} \label{equ9}
\left[\rho^{(i_Aj_A)},\left(\rho^{(i_Aj_A)}\right)^{\dag}\right]=0.
\end{align}
The matrix satisfying Eq. (\ref{equ8}) or (\ref{equ9}) is called
normal matrix\cite{15}. In addition, since all $\rho^{(i_Aj_A)}$ are
diagonalized by the same unitary matrix $U$,
\begin{align} \label{equ10}
\rho^{(i_Aj_A)}=U\Lambda ^{(i_Aj_A)}U^{\dag},
\end{align}
they have the same eigen-vectors. Any two normal matrices have the
same eigenvectors if and only if they commute with each
other\cite{15}. Consequently, we can conclude the criterion for zero
quantum discord now: all $N^2$ blocks $\rho^{(i_Aj_A)}$ in
Eq.(\ref{equ4}) are normal matrices (satisfying Eq.(\ref{equ9})),
and must commute with each other. This criterion has the advantage
that we can directly work on the matrix Eq.(\ref{equ2}) in any
tensor product basis, without the need to find out the particular
basis, $\{|k^{\prime}_B>\}$, required in the criterion of
Eq.(\ref{equ3}). As we all known, a bipartite state with zero
quantum discord, see Eq.(\ref{equ3}), must be a separable state.
Based on the new criterion, we can conclude that if a density matrix
is composed of diagonal block matrices, it represents a separable
state, which is valid for bipartite systems in any dimension.
However, inverse case is not true. A separable state does not
necessarily to have a density matrix composed of diagonal block
matrices. Given a high dimensional bipartite or multipartite quantum
state, how to efficiently verify its separability or entanglement is
still an open question.

We stress here that although commutation relations are used to
describe the criterion of zero quantum discord, just as done in
ref.\cite{14}, the present criterion has no direct connection with
that in ref.\cite{14}. The number of commuting matrices used in
ref.\cite{14}, denoted as $L$, is equal to the rank of the
correlation matrix, which is smaller than or equal to the minimal
one between the two squared dimension degrees of the two subsystems
A and B, i.e., $L\leq \min\{N^2,M^2\}$. However, The number of
commuting matrices used in our criterion is fixed as $N^2$, with $N$
being the dimension of the subsystem A. Furthermore, all the
commuting matrices used in ref.\cite{14} are Hermitian operators,
while the commuting matrices in our criterion can be non-Hermitian
or Hermitian, which depend on the density matrix itself.

If a quantum state $\rho_{AB}$ has been verified to have zero
quantum discord, we can obtain $U$ and $\{|k^{\prime}_B>\}$ by
diagonalizing any non-zero one of the block matrices
$\rho^{(i_Aj_A)}$ in Eq.(\ref{equ5}). From Eqs. (\ref{equ6}) and
(\ref{equ10}), we have
\begin{align} \label{equ13}
\rho_{B}=\text{Tr}_A(\rho_{AB})=\sum_{i=1}^N\rho^{(i_Ai_A)}=\sum_{i=1}^NU\Lambda^{(i_Ai_A)}U^{\dag}=UDU^{\dag},
\end{align}
with the diagonal matrix $D=\sum_{i=1}^N\Lambda^{(i_Ai_A)}$, which
tells us that the reduced matrix $\rho_B$ can also be diagonalized
by the unitary matrix $U$.

Let us now consider an example. Given a class of two-qubit X-state
in the basis of $\{|1_A1_B>,|1_A2_B>,|2_A1_B>,|2_A2_B>\}$,
\begin{align} \label{equ14}
\rho_x=\left(
         \begin{array}{cccc}
           x & 0 & 0 & \sqrt{x(0.5-x)} \\
           0 & 0.5-x & \sqrt{x(0.5-x)} & 0 \\
           0 & \sqrt{x(0.5-x)} & x & 0 \\
           \sqrt{x(0.5-x)} & 0 & 0 & 0.5-x \\
         \end{array}
       \right), ~~~\left(x\in[0,0.5]\right),
\end{align}
we can check whether the above states have zero quantum discord through three steps.\\
(1) Partition the density matrix (\ref{equ14}) into four blocks:
\begin{align} \label{equ15}
\rho^{(1_A1_A)}=\rho^{(2_A2_A)}=\left(
         \begin{array}{cc}
           x & 0  \\
           0 & 0.5-x \\
         \end{array}
       \right), ~~~\rho^{(1_A2_A)}=\rho^{(2_A1_A)}=\left(
         \begin{array}{cc}
           0 & \sqrt{x(0.5-x)}  \\
           \sqrt{x(0.5-x)} & 0 \\
         \end{array}
       \right).
\end{align}\\
(2) Check whether the four blocks are normal matrices (satisfying
Eq.(\ref{equ9})): Yes here.\\
(3) Check whether all of them commute with each other: as
\begin{subequations}
\begin{align} \label{equ15a1}
\rho^{(1_A1_A)}\rho^{(1_A2_A)}=\left(
        \begin{array}{cc}
           0 & x\sqrt{x(0.5-x)}  \\
           (0.5-x)\sqrt{x(0.5-x)} & 0 \\
         \end{array}
       \right)
\end{align}\\
      \text{and}
       \begin{align} \label{equ15a2}
       \rho^{(1_A2_A)}\rho^{(1_A1_A)}=\left(
        \begin{array}{cc}
           0 & (0.5-x)\sqrt{x(0.5-x)}  \\
           x\sqrt{x(0.5-x)} & 0 \\
         \end{array}
       \right),
       \end{align}\\
       \end{subequations}
the equality
$\rho^{(1_A1_A)}\rho^{(1_A2_A)}=\rho^{(1_A2_A)}\rho^{(1_A1_A)}$
holds true only when $x=0$, $0.25$ or $0.5$. In the case of
$x=0.25$, the unitary transformation $U=\frac{\sqrt{2}}{2}\left(
        \begin{array}{cc}
           1 & 1  \\
           1 & -1 \\
         \end{array}
       \right)$ diagonalizes the four matrices in Eq.(\ref{equ15}), and the local projectors $\{|k^{\prime}_B>\}$ in the pointer state are $|1^{\prime}_B>=\frac{\sqrt{2}}{2}(|1_B>+|2_B>)$ and
$|2^{\prime}_B>=\frac{\sqrt{2}}{2}(|1_B>-|2_B>)$. For $x=0$ and
$0.5$, the four matrices in Eq.(\ref{equ15}) are already diagonal(or
zero matrix), and $\{|1_B>,|2_B>\}$ is just the set of local
projectors used in the pointer state.

The quantum discord of this state can directly be calculated by
using the results in
refs.\cite{11,11add1,11add2,11add3,11add4,11add5}, which is
\begin{align} \label{equ16}
\delta(\rho_x)=&-1-(2x)\text{Log}_2(2x)-(1-2x)\text{Log}_2(1-2x)\nonumber \\
&-\sum_{k=1}^2[0.5+(-1)^k\sqrt{2x(1-2x)}]\text{Log}_2[0.5+(-1)^k\sqrt{2x(1-2x)}].
\end{align}
The zero quantum discord occurs only when $x=0$, $0.25$ or $0.5$,
which is the same as predicted by using our criterion.

\section{Proposed experiment and discussions}
Now we propose an experimental scheme to test the criterion based on
the above X-state (\ref{equ14}), which can be generated through the
following procedure. The entangled photon pairs from the type-I
parametric down conversion are in the state,
\begin{align} \label{equ17}
|\psi_1>=\text{cos}\theta|H_AH_B>+\text{sin}\theta|V_AV_B>,
\end{align}
where $\theta$ is the angle of the pump polarization direction with
respect to the vertical orientation, and the two optical axis of the
non-linear crystals are arranged in horizontal and vertical
orientations (H and V)\cite{16,17}, respectively, see Fig. 1. An
electro-optical modulator (EOM) in path A, switched on (acting as a
half-wave plate) or off (performing nothing) through the control of
a random number generator (RNG), convert, with probability $50\%$,
the polarization of the photon in path A from V to H, or vice
versus\cite{18}. The quantum state after the EOM is,
\begin{subequations} \label{equ18}
\begin{align} \label{equ18a}
\rho_{AB}=0.5|\psi_1><\psi_1|+0.5|\psi_2><\psi_2|,
\end{align}
\text{with}
\begin{align} \label{equ18b}
|\psi_2>=\text{cos}\theta|V_AH_B>+\text{sin}\theta|H_AV_B>.
\end{align}
\end{subequations}
The density matrix of the above state $\rho_{AB}$ in the basis
$\{|H_AH_B>,|H_AV_B>,|V_AH_B>,|V_AV_B>\}$ is just the X-state
(\ref{equ14}) with $x=0.5\text{cos}^2\theta$, which can be
experimentally measured through the two-qubit tomography\cite{19}.
Now we have all the four block matrices, and we can apply them in
our criterion to tell whether the quantum discord is zero or not. In
order to verify the zero quantum discord for $x=0.25$ and non-zero
quantum discord for $x\neq 0,0.25,0.5$ experimentally, we use the
following procedure. A half wave plate (HWP) in path B rotates the
polarization of the photon in this path by the angle of $45^0$,
which corresponds to the unitary transformation
$U=\frac{\sqrt{2}}{2}\left(
        \begin{array}{cc}
           1 & 1  \\
           1 & -1 \\
         \end{array}
       \right)$, as mentioned above. The H and V photons in this path will be
separated by the polarization beam splitter (PBS) and then register
on the two detectors $D_1$ and $D_2$, respectively, which
corresponds to two local orthogonal projectors on the photon B. Once
the photon B is detected by $D_1$ (with probability $p_1=2x$), or by
$D_2$ (with probability $p_2=1-2x$), the other photon in path A will
turn to the state $\rho^A_1$ or $\rho_2^A$, accordingly, which can
be found out through the single-qubit tomography\cite{19}. With the
measured $\rho^A_1$ and $\rho_2^A$, we can construct a density
matrix for the AB system,
\begin{align} \label{equ19}
\rho_{measure}=p_1\rho^A_1\otimes(|\phi_1><\phi_1|)+p_2\rho^A_2\otimes(|\phi_2><\phi_2|),
\end{align}
where $|\phi_1>=\frac{\sqrt{2}}{2}(|H_B>+|V_B>)$ and
$|\phi_2>=\frac{\sqrt{2}}{2}(|H_B>-|V_B>)$ are the two eigenvectors
associated with the measurement in current experimental setup. For
$x=0.25$, Eq.(\ref{equ19}) is equal to equation (\ref{equ14}), which
means zero quantum discord. For $x\neq 0,0.25,0.5$, Eq.(\ref{equ19})
will not be equal to the state of equation (\ref{equ14}), no matter
what kind of local projectors on the photon B are chosen, which
means non-zero quantum discord. For the two trivial cases of $x=0$
or $0.5$, the zero quantum correlation of the state (\ref{equ14})
can be verified via the same method as above by removing the HWP in
path B.

A useful result can be derived from our criterion: any POVM can't
extinguish the quantum correlation of the bipartite state with
nonzero quantum discord. For a bipartite quantum state $\rho_{AB}$,
we do a POVM on subsystem B by attaching an ancillary system
$\rho_{C}$ on it, and making a projective measurement on the
extended BC system. The new bipartite state, composed of one
subsystem A and another subsystem B plus C, is $\rho_{A(BC)}=
\rho_{AB}\otimes \rho_{C}$, which has zero quantum discord if and
only if the quantum discord of the original bipartite quantum state
$\rho_{AB}$ is zero, no matter what kind of ancillary system is
chosen. The matrix product rule\cite{20} directly gives us the
following equality,
\begin{align} \label{equ17}
\left[\rho^{(i_Aj_A)}\otimes
\rho_{C},\rho^{(i^{\prime}_Aj^{\prime}_A)}\otimes \rho_{C}\right]=
\left[\rho^{(i_Aj_A)},\rho^{(i^{\prime}_Aj^{\prime}_A)} \right]
\otimes (\rho_{C}\rho_{C})
\end{align}
and thus the above statement can be easily proven through the
commutation relations among the blocks of the original density
matrix $\rho_{AB}$, and also of the density matrix $\rho_{A(BC)}$.
Therefore, the above criterion for zero quantum discord is valid for
all types of local measurement, including POVM.

\section{conclusions}
To summarize, we derive a new criterion for zero quantum discord of
arbitrary bipartite states, which is easy to be implemented with
three steps: (1) Partition the density matrix of the $N\otimes M$
quantum state into $N^2$ block matrices; (2) Check whether every
block is a normal matrix (commuting with its Hermitian transpose);
(3) Check whether all block matrices commute with each other. For a
bipartite state with zero quantum discord, we can find out the set
of projectors, which do not change the whole state after being
imposed on one of the subsystems, by diagonalizing any non-zero
block matrix of its density matrix. This set of projectors provides
a way to locally access the information in the system without
disturbing the whole state. A class of two-qubit X-state is used to
test the criterion, which can be experimentally implemented. It is
also shown that POVM can't extinguish the quantum correlation of a
bipartite state with nonzero quantum discord, although it may have
an effect on the evaluation of non-zero quantum correlation.

\section{Acknowledgments}
We wish to thank Prof. J.P. Dowling for discussions and improving
the English. This research was supported by national natural science
foundation of China (NSFC) under Grant No. (10804042), the National
Basic Research Program of China (2011CB922203), RGC Grant No.
(HKBU202910) of HK Government and FRG of HK Baptist University.

\newpage

\begin{figure}
\begin{center}
\includegraphics[width=6.2in]{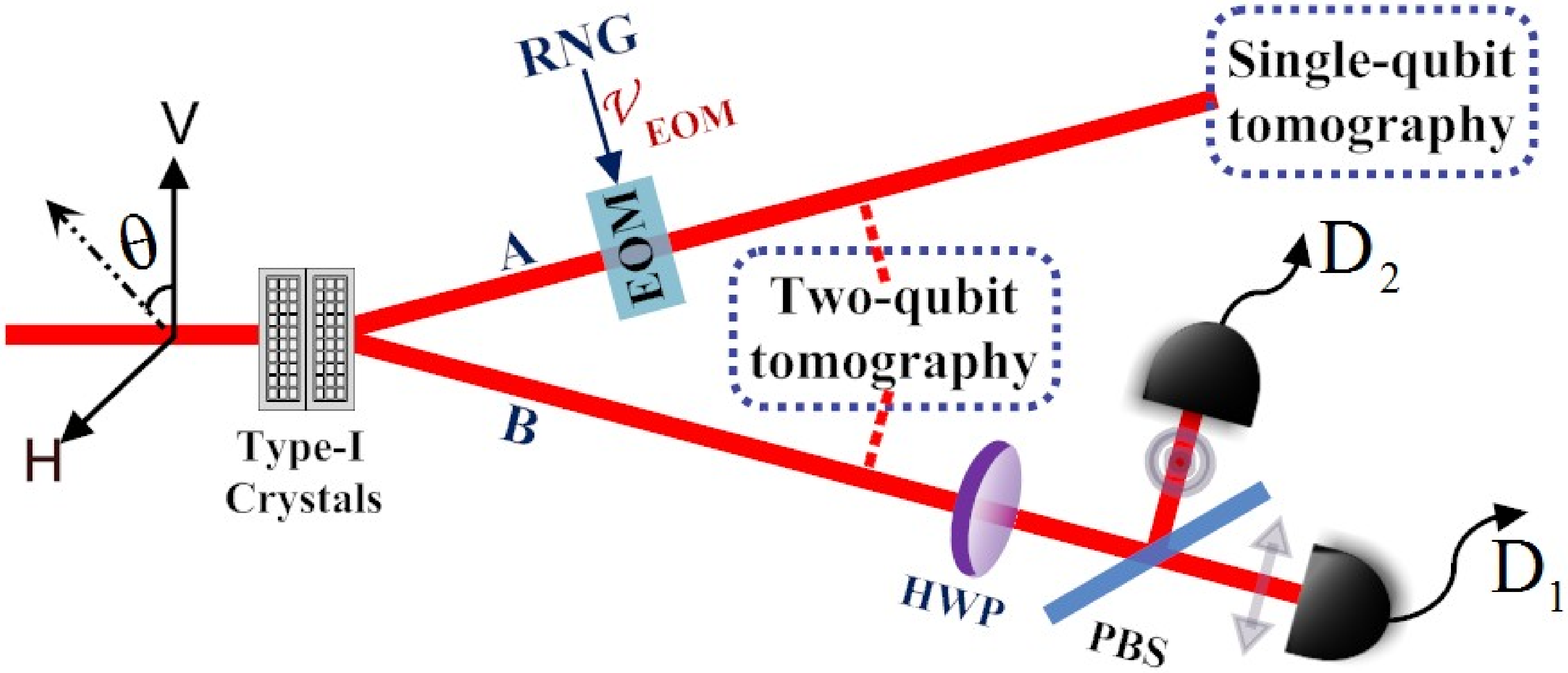}
\end{center}
\caption{Proposed experimental setup: one of the two entangled
photons (B), produced from non-linear crystals via type-I parametric
down conversion, is sent to two single-photon detectors, $D_1$ and
$D_2$, after its polarization is rotated by a half wave plate (HWP).
The polarization beam splitter (PBS) is used to distinguish the two
types (H or V) of polarization of the photon B, and separate them.
The polarization of the other photon (A) is inverted, with
probability $50\%$, by the electro-optical modulator (EOM) and then
measured through single-qubit tomography. The quantum state of the
two photons after the EOM can be measured through two-qubit
tomography.}\label{fig1}
\end{figure}

\end{document}